\documentclass[conference]{IEEEtran}
\IEEEoverridecommandlockouts
\usepackage{cite}
\usepackage{amsmath,amssymb,amsfonts}
\usepackage{algorithmic}
\usepackage{graphicx}
\usepackage{booktabs}
\usepackage{textcomp}
\usepackage{xcolor}
\def\BibTeX{{\rm B\kern-.05em{\sc i\kern-.025em b}\kern-.08em
    T\kern-.1667em\lower.7ex\hbox{E}\kern-.125emX}}

\newcommand{\add}[1]{\textcolor{black}{#1}}
    
\begin{document}

\title{SOUL-Net: A Sparse and Low-Rank Unrolling Network for Spectral CT Image Reconstruction\\
\thanks{X. Chen, W. Xia, Z. Yang, H. Chen, Y. Liu and J. Zhou are with the College of Computer Science, Sichuan
University, Chengdu 610065, China (e-mail: xchen6944@gmail.com; xwj90620@gmail.com; cziyuanyang@gmail.com; huchen@scu.edu.cn; liuyan77@scu.edu.cn; zhoujl@scu.edu.cn).}
\thanks{
Y. Zhang is with the School of Cyber Science and Engineering, Sichuan University,
Chengdu 610065, China, and with the College of Computer Science, Sichuan
University, Chengdu 610065, China (e-mail: yzhang@scu.edu.cn). (Corresponding
author: Y. Zhang.)}
}
\author{Xiang Chen, Wenjun Xia, Ziyuan Yang, Hu Chen, Yan Liu, Jiliu Zhou,  Yi Zhang}

\maketitle

\begin{abstract}
    Spectral computed tomography (CT) is an emerging technology, that generates a multienergy attenuation map for the interior of an object and extends the traditional image volume into a 4D form. Compared with traditional CT based on energy-integrating detectors, spectral CT can make full use of spectral information, resulting in high resolution and providing accurate material quantification.
    Numerous model-based iterative reconstruction methods have been proposed for spectral CT reconstruction. However, these methods usually suffer from difficulties such as laborious parameter selection and expensive computational costs. In addition, due to the image similarity of different energy bins, spectral CT usually implies a strong low-rank prior, which has been widely adopted in current iterative reconstruction models. Singular value thresholding (SVT) is an effective algorithm to solve the low-rank constrained model. However, the SVT method requires manual selection of thresholds, which may lead to suboptimal results.
    To relieve these problems, in this paper, we propose a Sparse and lOw-rank UnroLling Network for spectral CT image reconstruction (SOUL-Net), that learns the parameters and thresholds in a data-driven manner.
   Furthermore, a Taylor expansion-based neural network backpropagation method is introduced to improve the numerical stability. 
    The qualitative and quantitative results demonstrate that the proposed method outperforms several representative state-of-the-art algorithms in terms of detail preservation and artifact reduction.

\end{abstract}

\begin{IEEEkeywords}
    Spectral CT, image reconstruction, deep learning, low-rank prior.
\end{IEEEkeywords}

\section{Introduction}
Due to the great potential for lesion detection and material decomposition, spectral CT has attracted increasing attention ~\cite{sct_intro_1, sct_intro_2, sct_intro_3} recently. By discriminating
photon energies during data acquisition, photon-counting detector (PCD)-based spectral CT can collect multiple sets of measured data with different spectral information within a single exposure. However, spectral projections acquired by PCD usually suffer from strong noise due to limited photons and the counting rate in each energy bin~\cite{intro_noise}. Therefore, the development of efficient spectral CT reconstruction algorithms is of great importance and urgently needed in clinical applications.

To date, dose-reduced scanning configuration has been widely used to alleviate the radiation risk. In this situation, directly reconstructing the spectral CT with traditional methods such as filtered back projection (FBP) and algebraic reconstruction technique (ART)~\cite{art} will result in severe artifacts and noise. To better suppress the noise and artifacts in spectral CT reconstruction, many researchers have been studying methods to improve the reconstruction quality of spectral CT. 

In recent years, a number of methods based on sparse regularization have been developed in CT reconstruction, including total variation (TV)~\cite{tv}, tight frames~\cite{tightFrame} and dictionary learning~\cite{dictlearn}. These methods can be extended to spectral CT directly by applying these sparsity constraints to each energy bin separately. However, such a strategy ignores the correlation among different energy bins, resulting in suboptimal results. 
As a result, many algorithms have been developed to make full use of the data correlation in the spectral domain~\cite{TDL, l0TDL, sct_spa}. For example, Zhang \textit{et al.} proposed a tensor dictionary learning (TDL) method for spectral CT reconstruction~\cite{TDL}, which effectively exploits the correlation among different energy bins.
To eliminate artifacts more effectively, Wu \textit{et al.} incorporated the $\ell_0$ norm into the TDL method ($\ell_0$TDL)~\cite{l0TDL}. Hu \textit{et al.} proposed a method termed spectral-image similarity-based tensor with enhanced-sparsity reconstruction (SISTER)~\cite{SISTER}.

On the other hand, low-rank is an important prior in spectral CT image reconstruction. This has been proven in numerous works~\cite{tensorNuclear, PRISM, tPRISM}. As an efficient constraint for low-rank regularization, the nuclear norm can be solved by the singular value thresholding (SVT) method~\cite{SVT}.  In~\cite{tensorNuclear}, Semerci \textit{et al.} proposed a method based on the tensor version of nuclear norm regularization. Wu \textit{et al.}~\cite{NLCTF} proposed a nonlocal low-rank cube-based tensor factorization (NLCTF) model for spectral CT reconstruction, that applies Kronecker-basis-representation (KBR) tensor factorization to the nonlocal similar spatial-spectral cubes to impose the low-rank constraint. 

Meanwhile, to enjoy the best of both worlds, a number of methods that integrate both sparsity and low-rank prior information have been proposed. Gao \textit{et al.} proposed a spectral CT
reconstruction method based on a prior rank, intensity and sparsity model (PRISM)~\cite{PRISM}. Li \textit{et al.} extended the PRISM into the tensor form (tPRISM)~\cite{tPRISM}. In~\cite{LRTDSTTV}, Li \textit{et al.} proposed a cerebral perfusion CT reconstruction (CPCT) model based on low-rank tensor Tucker decomposition and spatial-temporal total variation regularization (LRTD-STTV). Reference~\cite{NLSMD} presented an iterative reconstruction method for spectral CT based on nonlocal low-rank and sparse matrix decomposition (NLSMD). In~\cite{ASSIST}, the authors proposed a spectral CT reconstruction method aided by self-similarity in image-spectral tensors, that utilizes the self-similarity of patches in both spatial and spectral domains (ASSIST). In~\cite{FONTSIR}, Chen \textit{et al.} proposed a fourth-order nonlocal tensor sparse and low-rank decomposition model for spectral CT image reconstruction. These studies have demonstrated the merit that combining the sparse and low-rank priors is beneficial to the reconstruction performance. However, although these iterative methods are effective in improving the reconstruction results, they are usually computationally expensive, even for single-energy reconstruction~\cite{SISTER}. The processing of multispectral images inevitably brings extra time consumption that is unexpected in the clinic.
Another disadvantage of iterative methods is that there are usually several parameters that need to be determined manually~\cite{SISTER, scss}. Manual selection of these parameters is very labor-intensive and time-consuming and  may lead to a suboptimal result.

Recently, deep learning has achieved great success in different image processing tasks~\cite{ImgDenoOv, ImageSupSur, ImgSupSur}.
Inspired by these impressive results, a number of deep learning-based methods have been proposed for CT reconstruction~\cite{CtSurvey1, CtSurvey2, CtSurvey3}. When these methods are separately applied to each monochromatic slice, the similarities among different energy bins are neglected. In the field of deep learning based spectral CT reconstruction, studies have been scarce. A skip-encode U-net with $\ell_p^p$-loss and anisotropy total-variation (ULTRA) training framework were proposed for spectral CT~\cite{ULTRA}. Mustafa \textit{et al.} proposed a fast reconstruction model of sparse-view spectral CT based on the U-net architecture with multichannel input and output (DSIR)~\cite{DSIR}. 
However, these methods are all postprocessing-based networks for spectral CT, ignoring data consistency. Nor do these methods fully explore the prior information inherent in spectral CT images such as low rank and sparsity.

To solve the above problems, in this paper, we propose a Sparse and lOw-rank UnroLling Network for spectral CT image reconstruction (SOUL-Net). On the one hand, to better exploit the similarities among different energy bins, we impose a low-rank constraint on the reconstruction model and SVT is adopted to solve the low-rank constraint problem. Since SVT usually requires the manual selection of thresholds, which may lead to suboptimal results, a learnable threshold is embedded into SVT. On the other hand, backpropagation in the learned SVT may lead to numerical instability, so a Taylor expansion-based neural network backpropagation method is introduced to relieve this problem.
The qualitative and quantitative results demonstrate that the proposed method outperforms several representative state-of-the-art algorithms in terms of image detail preservation and artifact reduction.

The rest of this paper is organized as follows. Section II elaborates the proposed methods. Section III presents the experimental details and the results. Concluding remarks are provided in the last section.

\section{METHODOLOGY}
A general model for $\ell_1$ norm based regularized reconstruction is as follows:

\begin{equation}
    \min_\chi\frac{1}{2}\|A(\chi)-\zeta\|_F^2+\lambda \Vert \Psi(\chi) \Vert_1,\label{genModel}
\end{equation}
where $\chi\in \mathbb{R}^{N_h\times N_w\times N_s}$ denotes the spectral CT image with a spatial resolution of $N_h\times N_w$ and $N_s$ energy bins. $\zeta\in \mathbb{R}^{N_d\times N_v\times N_s}$ is the projection data acquired with a scanning geometry of $N_d$ detector elements and $N_v$ projection views. $A\in \mathbb{R}^{(N_d\cdot N_v)\times( N_h\cdot N_w)}$ is the system matrix for the specific scanning geometry and $A(\cdot)$ can be implemented by the multienergy version of the Radon transform. $\mathrm{\Psi}$ denotes the sparsifying transform for $\chi$. $\|\cdot\|_F$ and $\|\cdot\|_1$ denote the Frobenius norm and $\ell_1$ norm, respectively. $\lambda$ is the regularization parameter.

As mentioned above, many works~\cite{tv, TDL} have been proposed based on the low-rank regularization terms and have shown the effectiveness of the low-rank prior in spectral CT reconstruction. To exploit the merits of both low-rank prior and sparse regularization,
Eq. (\ref{genModel}) can be extended as

\begin{equation}
    \min_\chi\frac{1}{2}\|A(\chi)-\zeta\|_F^2+\lambda_{1} \Vert \mathrm{\phi}(\chi) \Vert_*+\lambda_2 \Vert \Psi(\chi) \Vert_1,\label{model}
\end{equation}
where $\lambda_1$ and $\lambda_2$ are the weighting parameters for the low-rank and sparse regularization terms, respectively. $\mathrm{\phi}$ denotes the operator that reshapes the tensor $\chi$ to a matrix with a size of $\mathbb{R}^{(N_h \cdot N_w)\times N_s}$. $\Vert \cdot \Vert_{*}$ represents the nuclear norm. 

 To simplify the optimization, $\mathrm{\phi}(\chi)$ can be replaced with an auxiliary variable $\Gamma$. Then, Eq. (\ref{model}) can be rewritten as

\begin{equation}
        \min_\chi\frac{1}{2}\|A(\chi)-\zeta\|_F^2+\lambda_{1} \Vert \Gamma \Vert_*+\lambda_2 \Vert \Psi(\chi) \Vert_1, \ \mathrm{s.t.}\,  \Gamma=\mathrm{\phi}(\chi). \label{2}
\end{equation}
The augmented Lagrange form of Eq. (\ref{2}) is formulated as
\begin{figure*}
    \centering
    \includegraphics[width=1.0\textwidth]{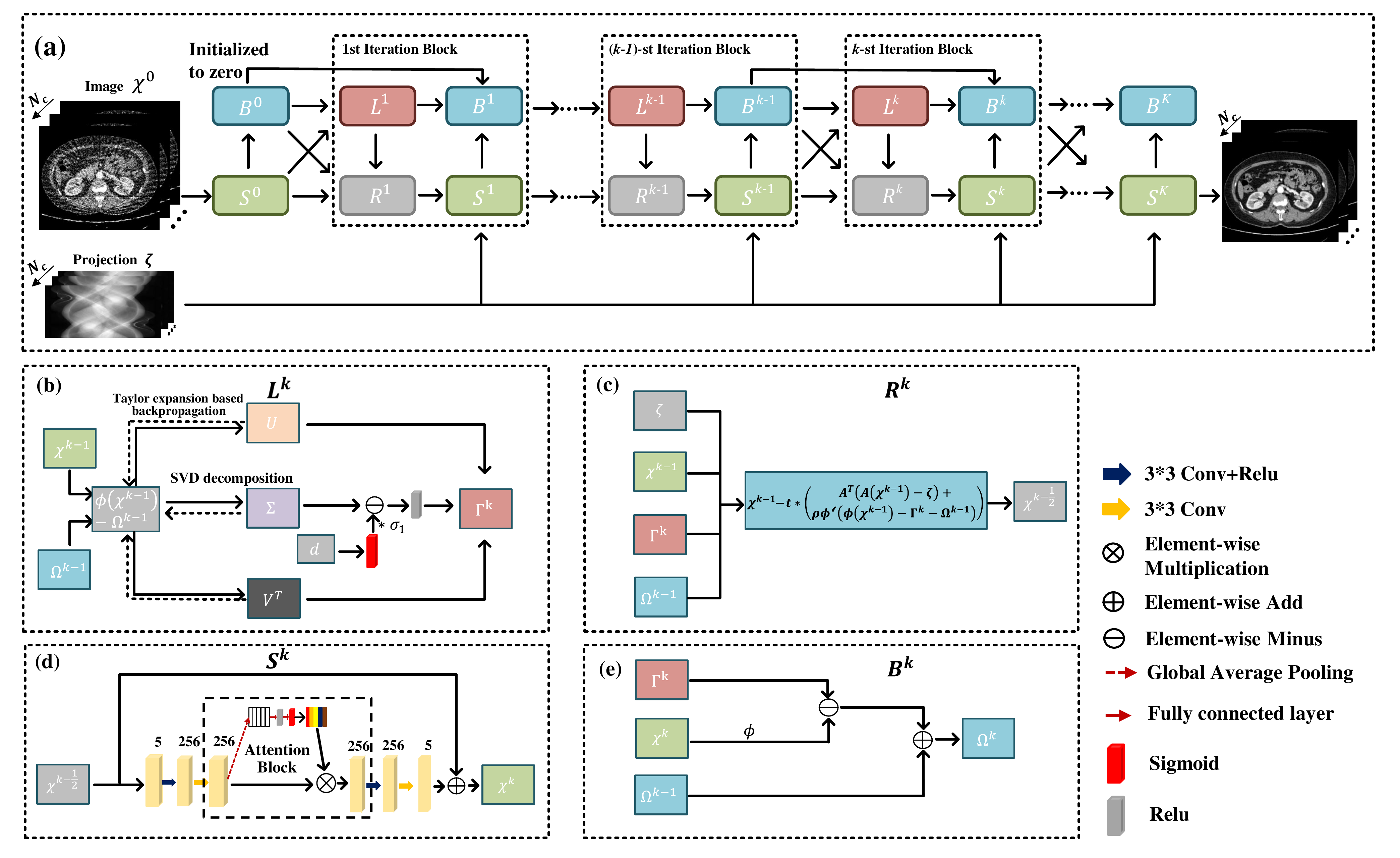}
    \caption{Overall structure of our proposed SOUL-Net network. SOUL-Net is defined over iterative procedures of Eq. \ref{subproblems}. Subfigure (a) shows optimization procedure for K iterative modules of SOUL-Net method. Four modules $L^k$,$R^k$, $S^k$ and $B^k$ in black dashed box correspond to four equations of Eq. \ref{subproblems}. Subfigure (b) represents $L^k$ module of each iteration block. Subfigure (c) represents $R^k$ module of each iteration block. Subfigure (d) corresponds to $S^k$ module of each iteration block. Subfigure (e) corresponds to $B^k$ module of each iteration block. $N_c$ is number of energy bins.}
    \label{fig:method}
\end{figure*}
\begin{equation}
    \begin{matrix}\min_\chi&\frac{1}{2}\|A(\chi)-\zeta\|_F^2+\lambda_{1} \Vert \Gamma \Vert_*+\lambda_2 \Vert \Psi(\chi) \Vert_1 \\&
    +<\alpha, \Gamma-\mathrm{\phi}(\chi)>+\frac{\rho}{2}\Vert \Gamma-\mathrm{\phi}(\chi)\Vert_F^2\end{matrix} ,\label{4}
\end{equation}
where $\alpha$ is the Lagrangian multiplier, and $\rho$ is the penalty factor. Let $\Omega=\alpha/\rho$. Then, a simplified version of Eq. (\ref{4}) can be obtained as

\begin{equation}
\begin{split}
        \min_\chi &\frac{1}{2}\|A(\chi)-\zeta\|_F^2+\lambda_{1} \Vert \Gamma \Vert_*+\lambda_2 \Vert \Psi(\chi) \Vert_1 \\
      &+\frac{\rho}{2}\Vert \Gamma-\mathrm{\phi}(\chi) +\Omega \Vert_F^2 \label{5}
\end{split}
\end{equation}
which can be iteratively solved using the alternating direction method of multipliers (ADMM) algorithm with scaled dual variables~\cite{ADMM} as follows:
\begin{subequations}
\begin{align}
 \Gamma^{k}=\arg\min_\Gamma& \frac{\rho}{2}\Vert \Gamma-\mathrm{\phi}(\chi^{k-1})+\Omega^{k-1}\Vert_F^2 + \lambda_{1} \Vert \Gamma \Vert_* , \label{sub_1}\\
 \chi^{k}=\arg\min_\chi &  \frac{1}{2}\|A(\chi)-\zeta\|_F^2+\lambda_2 \Vert \Psi (\chi) \Vert_1 + \notag\\
 & \frac{\rho}{2}\Vert \Gamma^{k}-\mathrm{\phi}(\chi)+\Omega^{k-1}\Vert_F^2, \label{sub_2}\\
 \Omega^{k}=\Omega^{k-1}+&\Gamma^{k}-\mathrm{\phi}(\chi^{k}).  \label{sub_3}
\end{align}
\end{subequations}

The subproblem in Eq. (\ref{sub_1}) is a nuclear norm minimization problem that can be solved with SVT~\cite{SVT}. We perform singular value decomposition as $(\phi(\chi^{k-1})-\Omega^{k-1})=U{\Sigma}V^T$. $U$ and $V^T$ denote the left singular matrix and right singular matrix of $(\phi(\chi^{k-1})-\Omega^{k-1})$ respectively. ${\Sigma}$ is the diagonal matrix of singular values of $(\phi(\chi^{k-1})-\Omega^{k-1})$. The result is obtained as
\begin{equation}
    \Gamma^{k}= {\Sigma}_i \max(\sigma_i-\lambda_1/\rho,0)u_i v_i^T
\end{equation}
where $\sigma_{i}$ is the $i$-th entry of the diagonal matrix $\Sigma$. $u_{i}$ and $v_{i}$ are the $i$-th column vectors of $U$ and $V$, respectively. 

Eq. (\ref{sub_2}) can be solved by using the iterative shrinkage-thresholding algorithm (ISTA) method~\cite{ISTA} as
\begin{subequations}
\begin{align}
\chi^{k-\frac{1}{2}}&= \chi-t\cdot( A^T (A(\chi^{k-1}) -\zeta)+ \notag\\
&\rho \mathrm{\phi}^{'}(\mathrm{\phi}(\chi^{k-1})-\Gamma^{k}-\Omega^{k-1})), \label{sub_2_split_1} \\
\chi^{k}=&\arg\min_\chi \frac{\rho}{2}\Vert \chi -\chi^{k-\frac{1}{2}}\Vert_F^2+ \lambda_2 \Vert \Psi \chi \Vert_1 \label{sub_2_split_2}  , 
\end{align} 
\end{subequations}
where $t$ denotes the step size, $\mathrm{\phi}^{'}$ denotes the conjugate operator of $\mathrm{\phi}$, and $A^T$ is the transpose of $A$.

The corresponding optimization procedure is summarized as follows:
\begin{subequations}
\begin{align}
L^k:\ &\Gamma^{k}= {\Sigma}_i \max(\sigma_i-\lambda_1/\rho,0)u_i v_i^T,   \label{subproblems_1}\\
R^k:\ &\chi^{k-\frac{1}{2}}= \chi^{k-1}-t\cdot( A^T (A(\chi^{k-1}) -\zeta)+ \notag\\
&\rho \phi^{’}(\phi(\chi^{k-1})-\Gamma^k-\Omega^{k-1})),  \label{subproblems_2}\\
S^k:\ &\chi^{k}=\arg\min_\chi \frac{\rho}{2}\Vert \chi -\chi^{k-\frac{1}{2}}\Vert_F^2+ \lambda_2 \Vert \Psi \chi \Vert_1,    \label{subproblems_3}\\
B^k:\ &\Omega^k=\Omega^{k-1}+\Gamma^{k}-\phi(\chi^k).  \label{subproblems_4} 
\end{align} \label{subproblems}
\end{subequations}

In our proposed SOUL-Net, the above iterative procedure is unrolled into a neural network, whose general architecture and components are illustrated in \add{Fig. \ref{fig:method}}. In the unrolled $k$-th iteration block shown in \add{Fig. \ref{fig:method}}(a), there are four main modules: low-rank module $L^k$, reconstruction module $R^k$, sparse module $S^k$ and Lagrangian multiplier module $B^k$. Each module corresponds to a subproblem in Eq. (\ref{subproblems}).  The structures of each module are illustrated in \add{Fig. \ref{fig:method}}(b)-(e), respectively. Each module is elaborated in the following subsections. 

The loss function of the proposed SOUL-Net is defined as
\begin{equation}
    \mathcal{L}=\frac{1}{N_s} \sum_{n=1}^{N_s}\Vert \widetilde \chi_n-\chi_n \Vert_1,
\end{equation}
where $\widetilde \chi_n$ denotes the label and $N_s$ is the total number of samples.

\subsection{Low-Rank Module}
The update of the low-rank module in Eq. (\ref{subproblems_1}) is illustrated in \add{Fig. \ref{fig:method}}(b). In the traditional SVT method, the selection of the threshold $\lambda_1/\rho$, which has a great impact on the final performance, is empirically set and time-consuming. To circumvent this obstacle, we introduce a Taylor expansion-based learnable threshold method ~\cite{RSVD} to solve the low-rank problem. This is shown as

\begin{equation}
    \Gamma= {\Sigma}_i \mathrm{Relu}(\sigma_i-\mathrm{Sigmoid}(d)*\sigma_1)u_i v_i^T
\label{lowrankpart}
\end{equation}
$\sigma_1$ is the largest singular value, and $d$ is the learnable parameter initialized to 0. The $\mathrm{Relu}$ activation function is essentially the $\mathrm{max}$ function in Eq. (\ref{subproblems_1}) and is used to guarantee the nonnegativity. $\mathrm{Sigmoid}(d)*\sigma_1$ is the learned threshold and has a range of $(0, \sigma_1)$. Eq. (\ref{lowrankpart}) can guarantees that the output is meaningful.

However, the backpropagation of singular value decomposition (SVD) may lead to numerical instability in the PyTorch framework~\cite{SVD_insta}. In an SVD-based network, we assume $X \in \mathbb{R}^{m\times n}=U{\Sigma}V^T$, $m>n$ and $\mathcal{L}$ is the loss function of the network. According to~\cite{MB}, only one of $\partial \mathcal{L}/\partial V$ and $\partial \mathcal{L}/\partial U^T$ is needed to complete the backpropagation, and the partial derivatives can be computed as
\begin{equation}
    \frac{\partial \mathcal{L}}{\partial X}=U\{2 \Sigma  ( K^T \cdot (V^T \frac{\partial \mathcal{L}}{\partial V} ))_{\mathrm{sym}} +(\frac{\partial \mathcal{L}}{\partial \Sigma})_{\mathrm{diag}} \}V^T
    \label{pari_L_X_1}
\end{equation}
and
\begin{equation}
     K_{i,j}=\left\{
    \begin{array}{ll}
        \frac{1}{\sigma_i^2-\sigma_j^2} , &  i \neq j\\
        0, &  i=j
    \end{array}
    \right.\label{k_i_j}
\end{equation}
where $\sigma_i$ and $\sigma_j$ are the $i$-th and $j$-th largest singular values. For a square matrix $M$, $M_{\mathrm{sym}}=(M^T+M)/2$ and $M_{\mathrm{diag}}$ is the resultant matrix by setting all off-diagonal elements of $M$ to 0.

However, if the distance between any two singular values is close to zero, then the partial derivative with respect to $X$ will be numerically unstable. In~\cite{RSVD}, the authors proposed a Taylor expansion-based method to compute the eigenvector gradients for covariance matrices. This achieves fast, accurate, and stable performance and can be easily embedded into a deep learning framework. 
As a result, to relieve the problem of SVT in gradient backpropagation, in this paper, we use a similar approach to calculate the gradient of SVD for a matrix. With the $Z$-th order Taylor expansion of $f(x)=\frac{1}{1-x}$, $x \in (0,1)$, $f(x)$ can be rewritten as 
\begin{equation}
    f(x)=1+x+...+x^Z+R_{Z+1},
\end{equation}
where $R_{Z+1}=\frac{x^{Z+1}}{1-x}$ is the remainder of the expansion. Eq. (\ref{k_i_j}) can be split into two cases: $\sigma_i>\sigma_j$ and $\sigma_i<\sigma_j$. For $\sigma_i>\sigma_j$, Eq. (\ref{k_i_j}) is converted into 

\begin{equation}
        K_{i,j}=\frac{1}{\sigma_i+\sigma_j}\frac{1}{\sigma_i}\frac{1}{1-\frac{\sigma_j}{\sigma_i}}\label{k_i_j_trans},\  \sigma_i>\sigma_j.
\end{equation}
The $Z$-th order Taylor expansion of Eq. (\ref{k_i_j_trans}) is formulated as
\begin{equation}
    K_{i,j}=\frac{1}{\sigma_i+\sigma_j}\frac{1}{\sigma_i}\left(1+\frac{\sigma_j}{\sigma_i}+...+(\frac{\sigma_j}{\sigma_i})^Z\right)+R_{Z+1},\  i<j,
\end{equation}
where
\begin{equation}
    R_{Z+1}=\frac{1}{\sigma_i^2-\sigma_j^2}(\frac{\sigma_j}{\sigma_i})^{(Z+1)},\ i<j.
\end{equation}

The solution for the case $\sigma_i<\sigma_j$ can be derived in the same manner. To avoid the numerical instability caused by the case $\sigma_i=0$, we add a small positive value $\epsilon$ to the diagonal of the matrix $X$, that is $X=X+\epsilon I$. This operation guarantees that $\sigma_i >=\sqrt{\epsilon}$ in the backpropagation of the neural network. More details can be found in~\cite{RSVD}. In the backpropagation of Eq. (\ref{subproblems_1}), we use this approach for threshold updating.

\subsection{Reconstruction Module}
\add{Fig. \ref{fig:method}}(c) illustrates the reconstruction module in Eq. (\ref{subproblems_2}).
Since the projection and backprojection operations are usually time-consuming, compute unified device architecture (CUDA) is adopted to accelerate the computation of Eq. (\ref{subproblems_2})~\cite{MAGIC}. To avoid the suboptimal results caused by manual selection of parameters, in this part, parameters $t$ and $\rho$ are learned from the training data.

\subsection{Sparse Module}
The update of sparse module in Eq. (\ref{subproblems_3}) is illustrated in \add{Fig. \ref{fig:method}}(d).
Eq. (\ref{subproblems_3}) is a standard $\ell_1$ norm problem that can be solved by a nonlinear method, and the solution is related to the specific form of the sparse transform. For instance, the solution of the widely used total variation (TV) norm problem~\cite{tv} can be obtained using the Chambolle-Pock projection algorithm~\cite{chambolle} as follows:

\begin{equation}
    \chi^k=\chi^{k-\frac{1}{2}} - \pi_{\lambda_2/\rho}(\chi^{k-\frac{1}{2}})
\end{equation}
where $\pi_{\lambda_2/\rho}(\cdot)$ is a nonlinear projection operator. To improve the performance, in our proposed SOUL-Net, an attention-based network is utilized to learn the nonlinear projection operator instead of the handcrafted sparse transform. As shown in the sparse module in \add{Fig. \ref{fig:method}}(d), $\chi$ is fed into a two-layer convolutional neural network (CNN). Then, the obtained feature maps are scaled with the learned channel attention~\cite{SEnet} and fed into another two-layer network. Finally, the output adds $\chi^{k-\frac{1}{2}}$ to complete the update of the sparse module. 

 The attention mechanism is widely used in medical imaging~\cite{MedAtt1, MedAtt2}, which identifies the parts of features that most contribute to the results. For spectral CT imaging, we introduce the channel attention mechanism ~\cite{SEnet} to better fuse the information across the energy bins and learn the nonlinear projection operator. The attention block is shown in \add{Fig. \ref{fig:method}}(d). The sizes of the convolution kernels are set to $3\times3$.

\subsection{Lagrangian Multiplier Module}

The update of the Lagrangian multiplier module in Eq. (\ref{subproblems_4}) is illustrated in \add{Fig. \ref{fig:method}}(e).

\begin{figure}[!htb]
    \centering
    \includegraphics[width=0.46\textwidth]{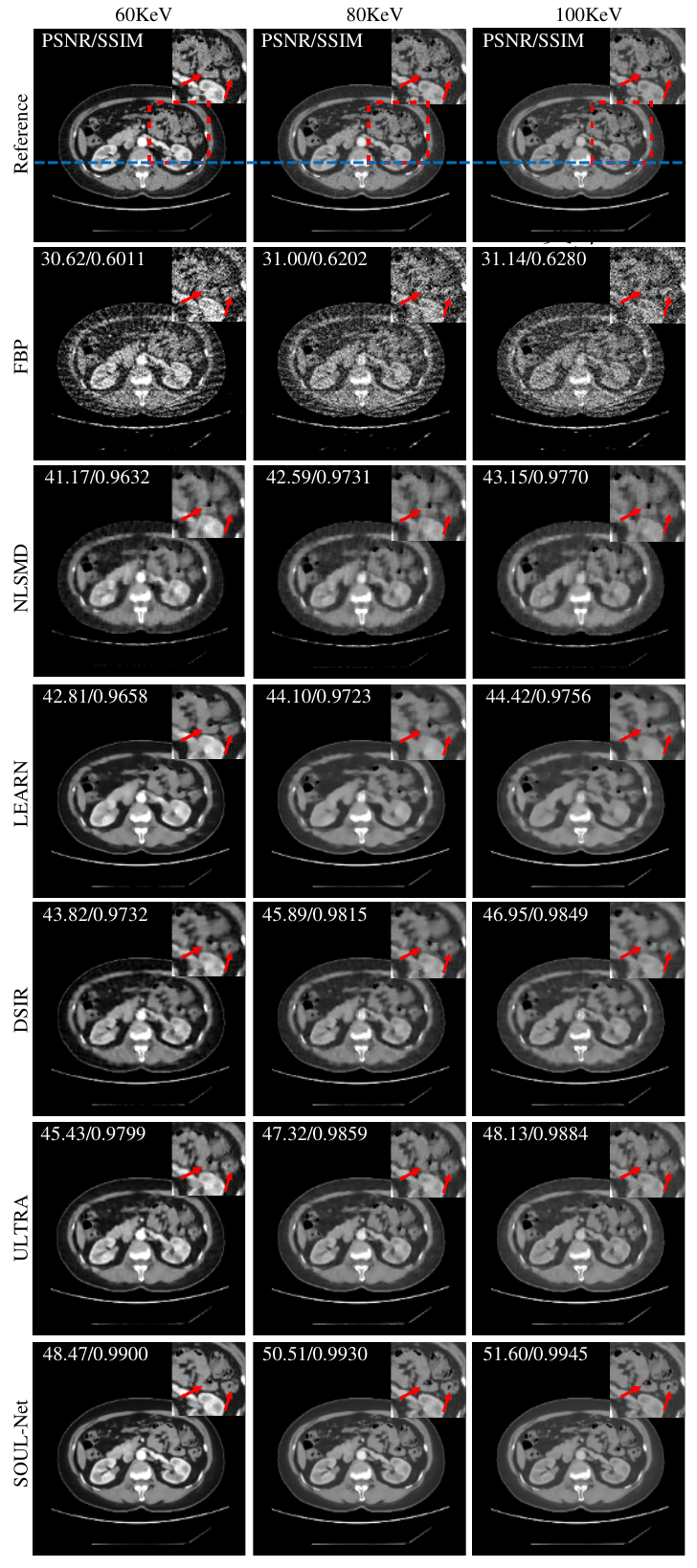}
    \caption{Reconstruction results of different methods with simulated noisy abdominal
data. Display window is [-160,240]HU.}
    \label{fig:fuqiang}
\end{figure}

\section{EXPERIMENTS AND RESULTS}
\subsection{Data Preparation}

To validate the performance of the proposed method, the
simulated clinical data is used in our experiments. The simulated clinical images are initially obtained using the GE Discovery dual-energy CT 750 HD scanner. In our experiments, 400 images are randomly selected from 4 patients as the training set, and 100 images from the remaining 1 patient are used as the test set. In total, virtual multichannel monochromatic images with a size of 256×256 from five different energy bins are produced using GE software from 60 to 100 keV in increments of 10 keV. The projection data are simulated with the distance-driven method with a fan beam geometry of~\cite{DisDri}. The distance between the scanner rotation center and X-ray source is set to 35 cm. The distance
between the radiation detector and scanner rotation center is set to
30 cm. The physical height and width of a pixel are both 0.72 mm. The detector has 1024 elements, each of which has a length of 0.58 mm. Sparse-view data are used to
verify the performance of the proposed SOUL-Net.
Sixty-four views evenly distributed over 360 views are chosen. In
addition, to verify the robustness of SOUL-Net in the clinical environment, Poisson and electronic noise are added to the simulated projection data according to~\cite{AddNoise} as follows:
\begin{equation}
    I=\mathcal{P}\left(I_0 \exp(-\zeta)\right)+\mathcal{N}(0, \sigma^2_e),
\end{equation}
where $I_0$ denotes the incident flux of the X-ray, $\sigma_e^2$ is the variance of the background electronic noise, and $\zeta$ represents the noise-free projection data. In our experiments, $I_0$ and $\sigma_e^2$ are set to $1.0\times{10}^6$ and 6, respectively.

\begin{figure}[!t]
    \centering
    \includegraphics[width=0.46\textwidth]{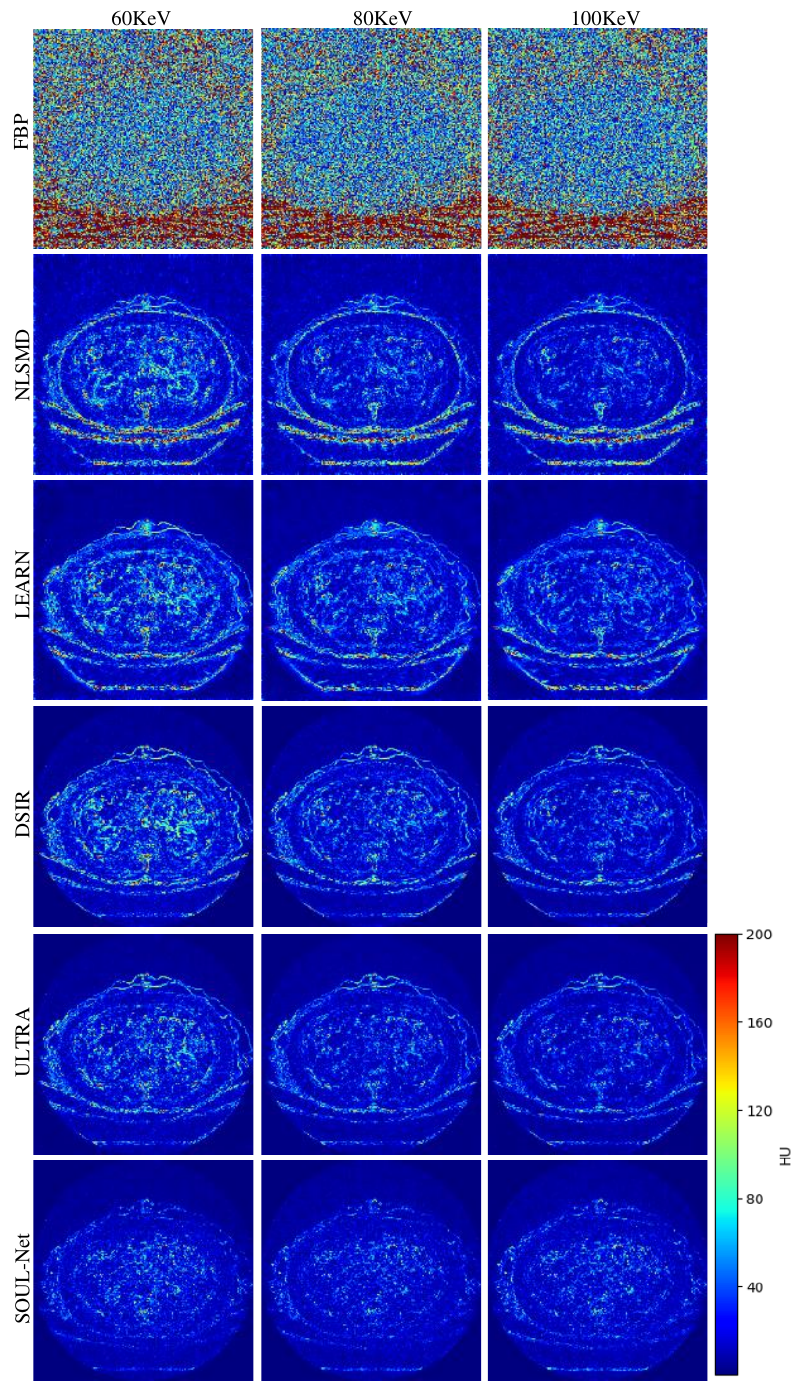}
    \caption{Absolute difference maps of abdominal images between different reconstruction results and references.}
    \label{fig:abso_fuqiang}
\end{figure}
\begin{figure*}[!htb]
    \centering
    \includegraphics[width=0.85\textwidth]{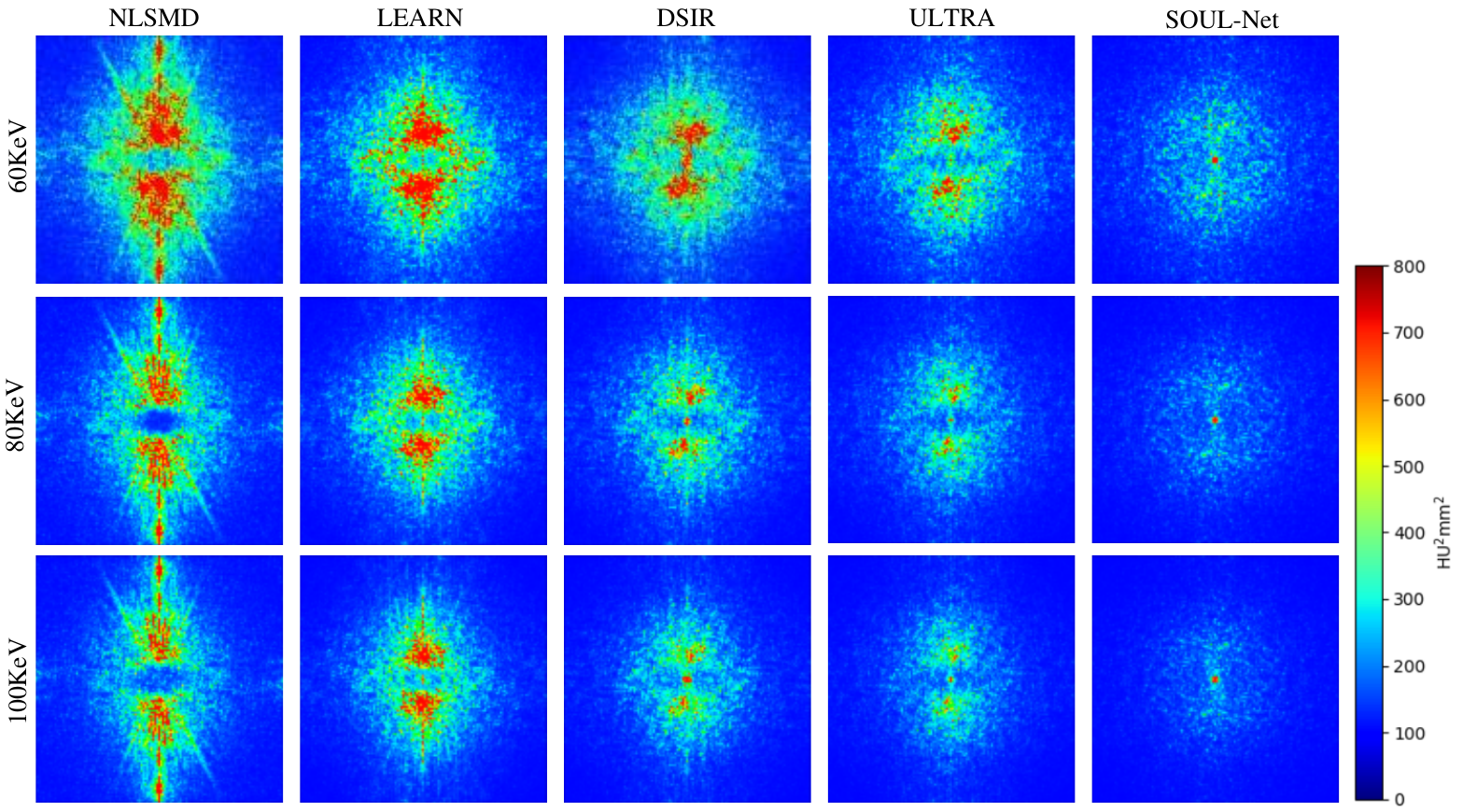}
    \caption{NPS maps of abdominal images from different reconstruction results in 60, 80 and 100 KeV energy bins.}
    \label{fig:fuqiang_NPS}
\end{figure*}

\begin{figure*}[!htb]
    \centering
    \includegraphics[width=1.0\textwidth]{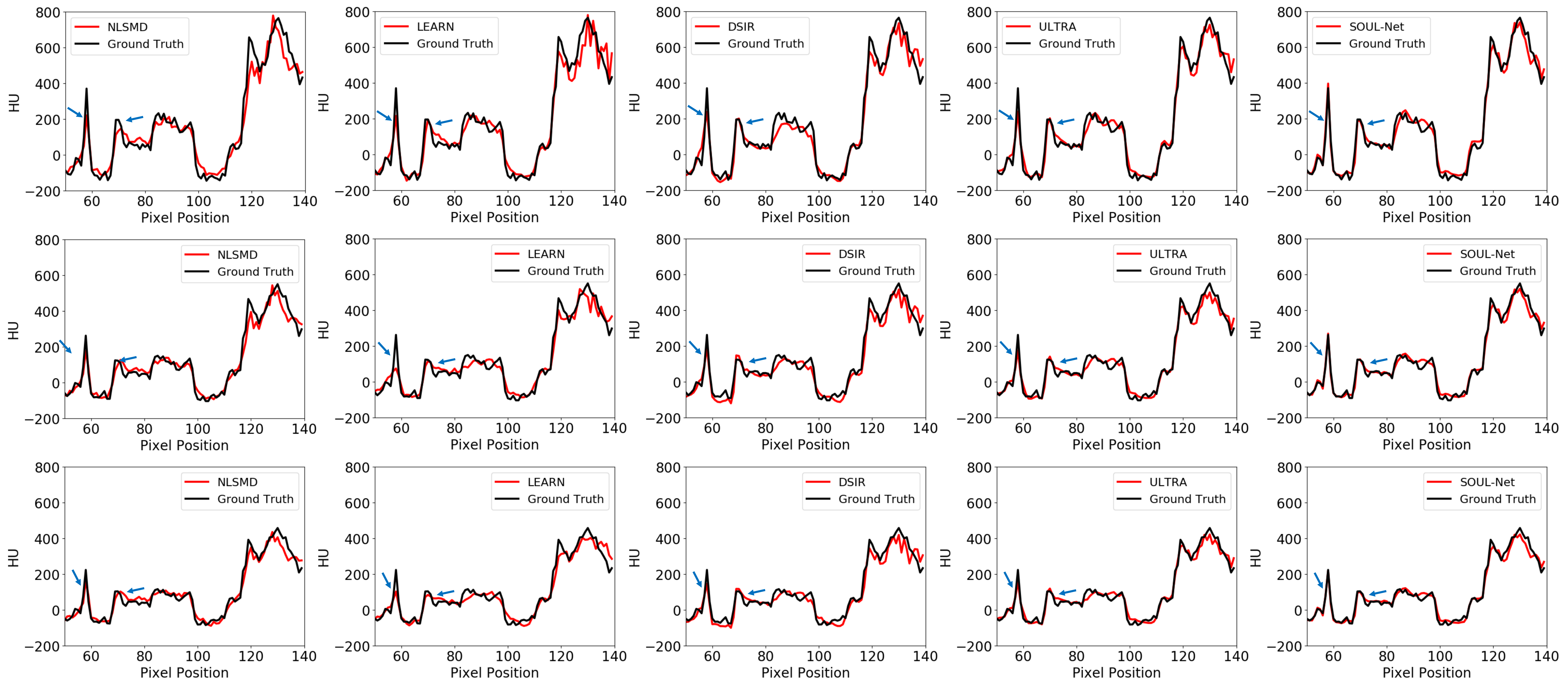}
    \caption{Horizontal profiles of abdominal results from different methods in 60, 80 and 100 keV energy bins from top to bottom. In columns from left to right, profiles come from NLSMD, LEARN, DSIR, ULTRA, and SOUL-Net.}
    \label{fig:fuqiang_profile}
\end{figure*}
\subsection{Experimental Configuration}

The experiments are performed in Python 3.9 with the PyTorch library on a PC (Intel Core i5 8400 CPU, 16 GB RAM and GTX 1080Ti GPU). The software code for this work is available at https://github.com/thesakura/SOUL-Net. In our experiments, the learning rate is set to $1\times10^{-4}$. The Adam solver is adopted to optimize our models with  ($\beta_1$,$ \beta_2$)=(0.9, 0.999)~\cite{Adam}. The batch size is set to 1. The number of iteration blocks is set to 10. $t$ is initialized to 0. $\rho$ is initialized to 1. $Z$ is set to 9. $\epsilon$ is set to $1\times10^{-8}$.

The peak signal-to-noise ratio (PSNR) and structural similarity index metric (SSIM) are employed to quantitatively evaluate the performance of different methods ~\cite{PSNRSSIM}. The PSNR is used to measure the error between the reconstructed spectral CT images and the reference spectral CT images, and is calculated as
\begin{equation}
    \mathrm{PSNR}=10\times\log_{10} \frac{\mathrm{MaxValue}^2}{\mathrm{MSE}},
\end{equation}
where MaxValue denotes the max value of the reference images, and MSE denotes the mean square error between the reconstructed images and reference images.
SSIM is adopted to measure the structural similarity between the reconstructed images and the reference images, and is defined as
\begin{equation}
    \mathrm{SSIM}=\frac{2e_Xe_{X^*}(\sigma_{XX^*}+c_2)}{(e_X^2+e_{X^*}^2+c_1)(e_X^2+e_{X*}^2+c_2)},
\end{equation}
where $X$ and $X^*$ represent the reconstructed image and reference image respectively. $e_X$ and $e_{X*}$ represent the mean values of $X$ and $X^*$ respectively. $\sigma_{X}$ and $\sigma_{X^*}$ denote the standard deviations of $X$ and $X^*$ respectively. $\sigma_{XX^*}$ represents the covariance between $X$ and $X^*$. $c_1$ and $c_2$ are two constants.
In addition, the noise power spectrum (NPS) is employed to evaluate the noise texture~\cite{NPS,NPS1}, which is defined as
\begin{equation}
    \mathrm{NPS}(f_i,f_j)=\frac{\Delta_i \Delta_j}{N_i N_j}\left<\vert \mathrm{DFT}({\Delta I(i,j)})\vert^2 \right>,
\end{equation}
where $\Delta_i$ and $\Delta_j$ are the physical sizes of the image pixels, and $N_i$ and $N_j$ denote the sizes of the selected region of interest (ROI) for the calculation of NPS. In this paper, $\Delta_i=\Delta_j=0.6641$mm and $N_i=N_j=127$. $\Delta I(i,j)$  represents the error component obtained by subtracting the reference image from the reconstructed image in the ROI. In~\cite{NPS}, the ROI is obtained from images with equal extraction intervals. In this paper, a total of $44\times44$ ROIs are chosen from the image of $256\times256$. The operator $<\cdot>$ denotes the average of all ROIs. $\mathrm{DFT}$ denotes the discrete Fourier transform.

In the experiments, four state-of-the-art methods NLSMD~\cite{NLSMD}, LEARN~\cite{LEARN}, DSIR~\cite{DSIR} and ULTRA~\cite{ULTRA} are used for comparison. NLSMD is a recently proposed iterative reconstruction
method for spectral CT based on nonlocal low-rank and sparse priors. LEARN is a classic unrolling network method for CT reconstruction. We reconstruct the images separately for each energy bin using the LEARN network in our experiments. DSIR and ULTRA are two postprocessing methods for spectral CT image reconstruction. The configuration and initialization of these competing methods are implemented according to the original literature or the recommendations.

\subsection{Results of Simulated Noisy Data}

\subsubsection{Results of Simulated Noisy Abdominal Image}

\add{Fig. \ref{fig:fuqiang}} shows one representative abdominal slice reconstructed with simulated noisy data using different methods. Three typical energy bins (60 keV, 80 keV and 100 keV) are displayed. It can be seen that some details in the NLSMD result are blurred,  although most noise and artifacts are suppressed. Similarly, some important structural details are smoothed by LEARN, DSIR and ULTRA. To better visualize the performance of detail recovery, the ROI indicated with red boxes in \add{Fig. \ref{fig:fuqiang}} are magnified. It can be observed that compared with other methods, SOUL-Net not only eliminates the noise and artifacts effectively but also maintains the structures effectively, especially in the parts indicated by red arrows. The SSIM and PSNR values of each reconstruction are shown in the upper left part of each subfigure in \add{Fig. \ref{fig:fuqiang}}. Our method achieves the highest SSIM and PSNR values.

\begin{figure}[!htb]
    \centering
    \includegraphics[width=0.46\textwidth]{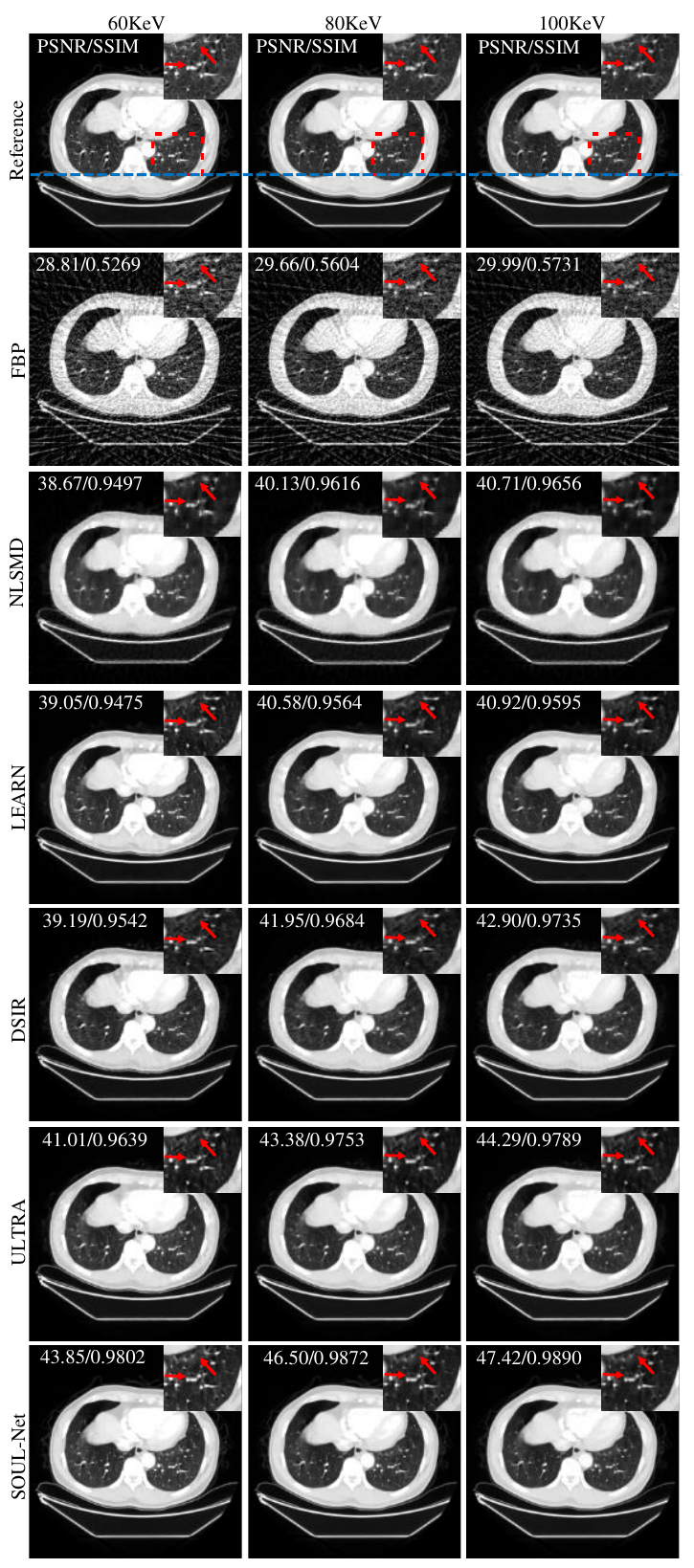}
    \caption{Reconstruction results for different methods with simulated noisy thoracic
data. Display window is [-1000,200]HU.}
    \label{fig:xiongqiang}
\end{figure}

\begin{figure}[!htb]
    \centering
    \includegraphics[width=0.46\textwidth]{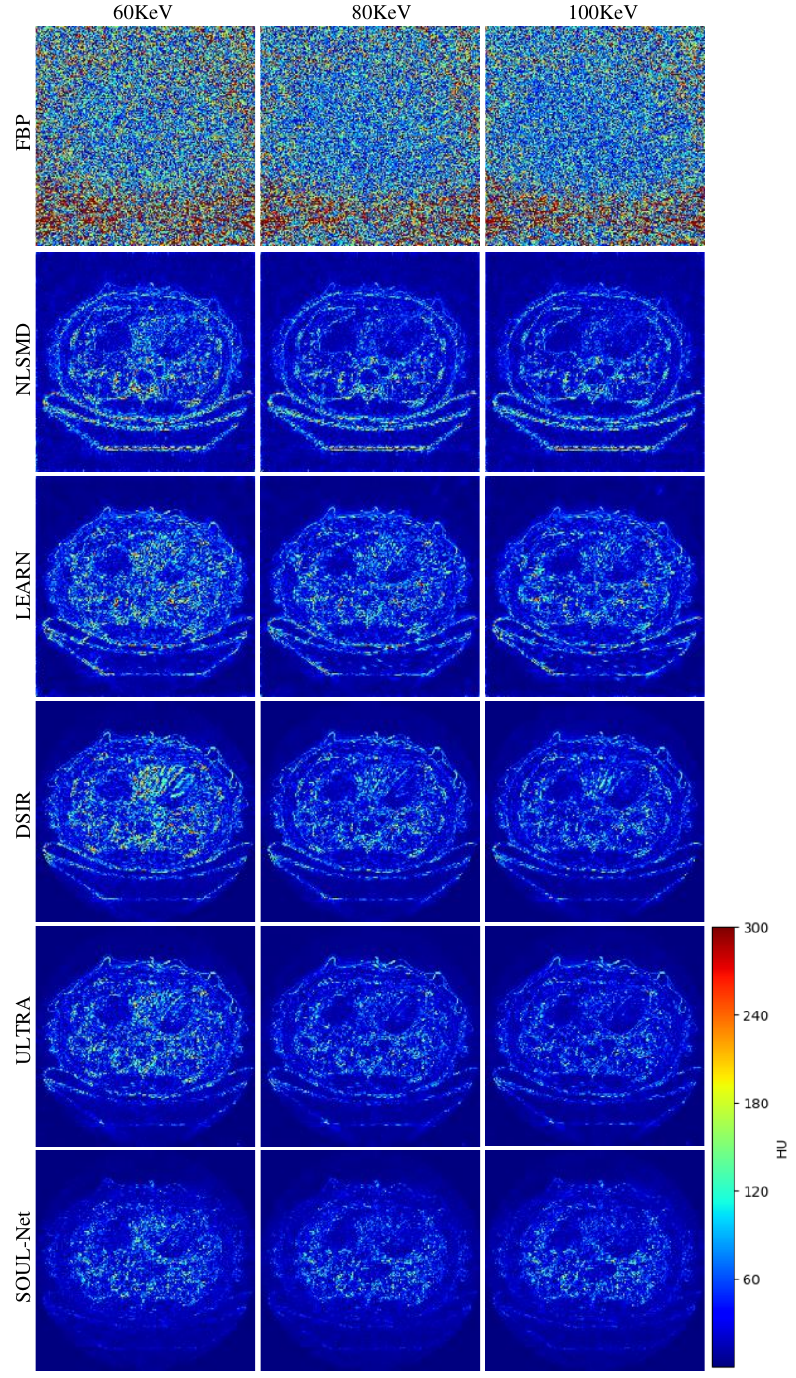}
    \caption{Absolute difference maps of thoracic images between reconstruction results and reference.}
    \label{fig:abso_xiongqiang}
\end{figure}
\begin{figure*}[!thb]
    \centering
    \includegraphics[width=0.85\textwidth]{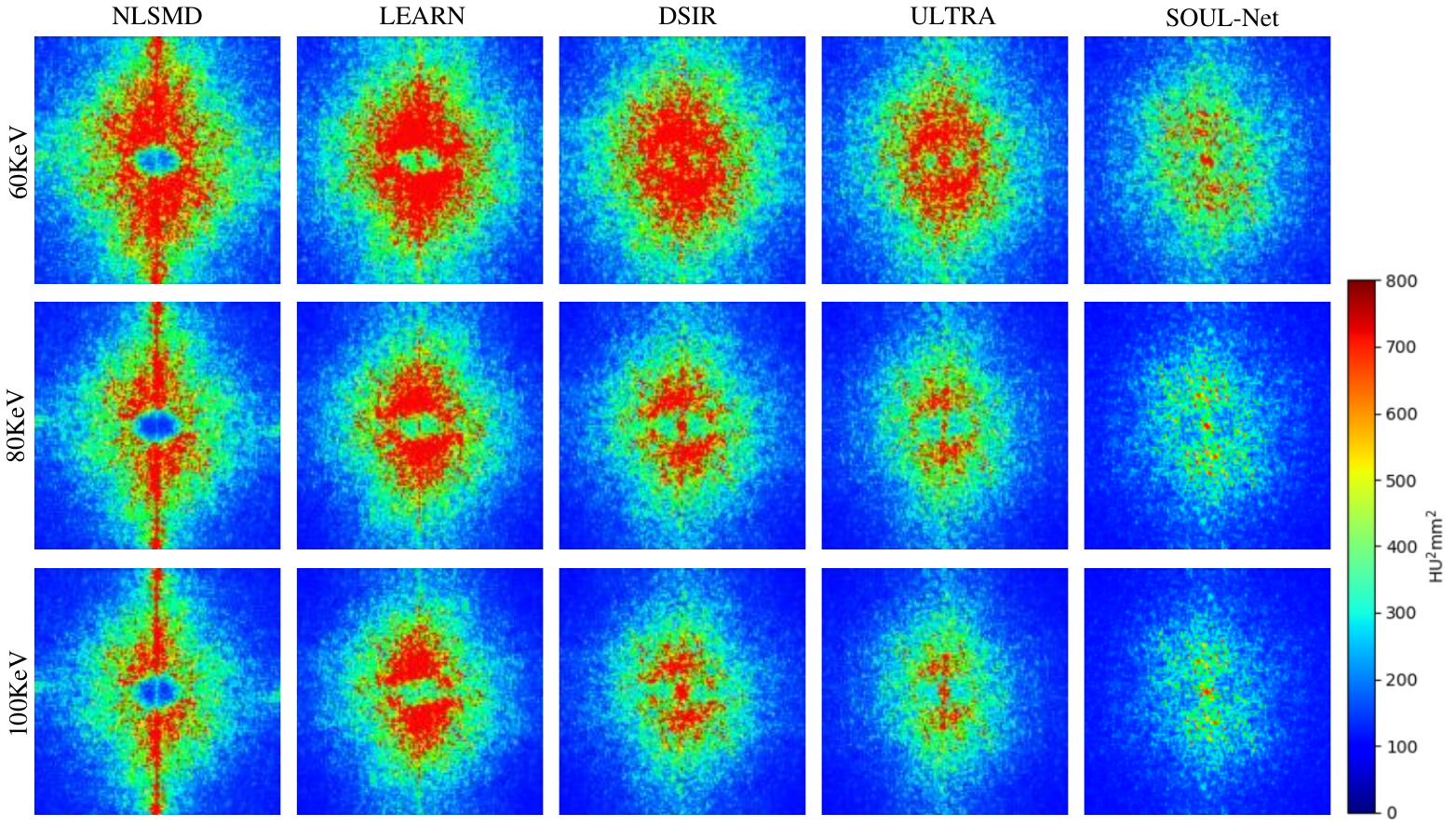}
    \caption{NPS maps of thoracic images from different reconstruction methods in 60, 80 and 100 KeV energy bins.}
    \label{fig:xiongqiang_NPS}
\end{figure*}

\begin{figure*}[!htb]
    \centering
    \includegraphics[width=1.0\textwidth]{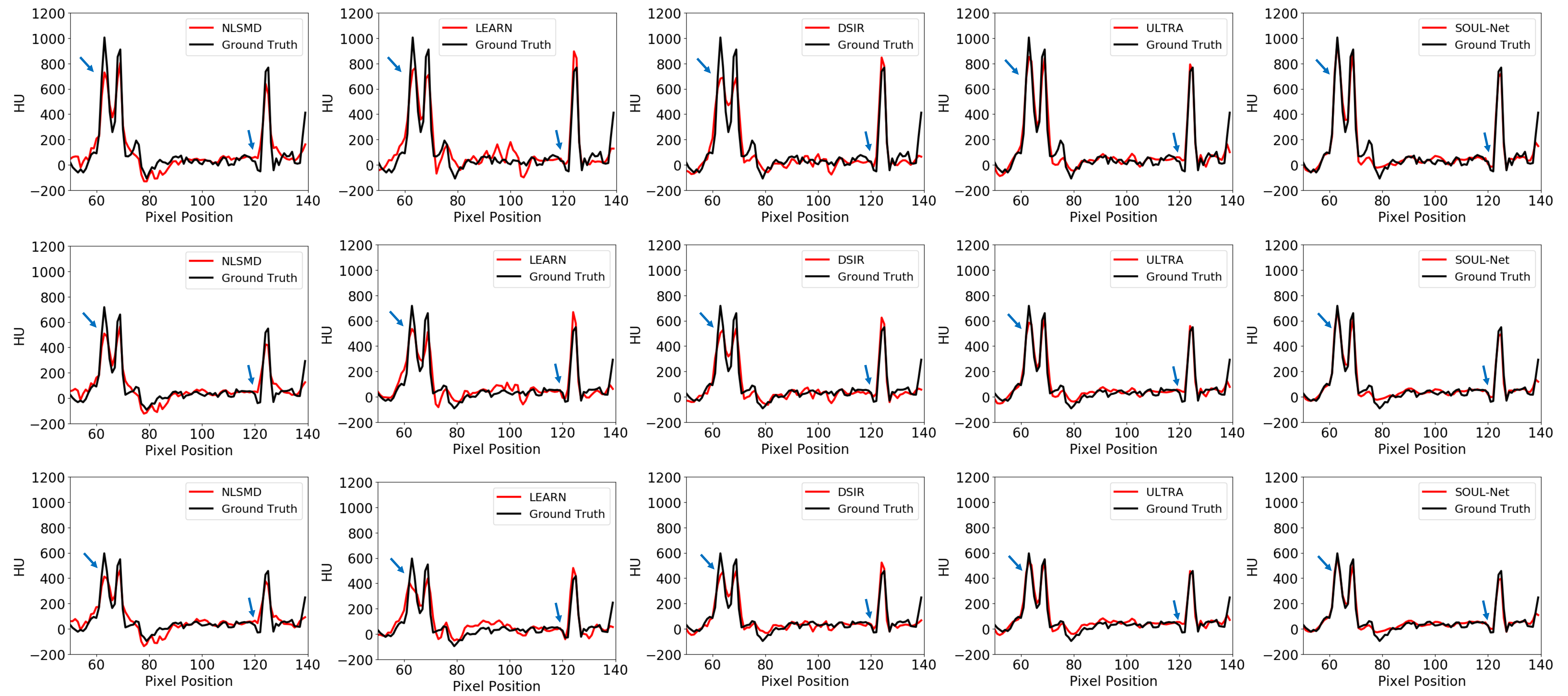}
    \caption{Horizontal profiles of thoracic results from different methods in 60, 80 and 100 keV energy bins from top to bottom. In columns from left to right, profiles come from NLSMD, LEARN, DSIR, ULTRA, and SOUL-Net.}
    \label{fig:xiongqiang_profile}
\end{figure*}

To better verify the detail recovery ability, \add{Fig. \ref{fig:abso_fuqiang}} shows the absolute difference images associated with the reference image. In \add{Fig. \ref{fig:abso_fuqiang}}, it is clear that SOUL-Net has the smallest residual compared with other methods. This indicates that our method is able to preserve the structural details effectively.

The NPS of all results are shown in \add{Fig. \ref{fig:fuqiang_NPS}}. We can see that in both high- and low-frequency bands, the error in the results of SOUL-Net is lower than that of the other methods. This can be treated as additional evidence of the effectiveness of our proposed method.

\add{Fig. \ref{fig:fuqiang_profile}} shows the horizontal profiles of different results along the blue dashed line in the first row of \add{Fig. \ref{fig:fuqiang}}. 
Two blue arrows indicate two sharp areas that can distinguish the ability in edge preservation for different methods . It is obvious that the proposed SOUL-Net has the closest results to the reference and achieves higher reconstruction accuracy than other methods. This is consistent with the results of our visual inspection.

\subsubsection{Results of Simulated Noisy Thoracic Image}

A typical thoracic case reconstructed using different methods is shown in \add{Fig. \ref{fig:xiongqiang}}. The lung window is used to show the details of the blood vessels. In the results of NLSMD, although noise and artifacts are well removed, most structural details in the lungs are blurred. LEARN and DSIR achieve better performance than NLSMD but still suffer from visible detail loss. ULTRA and SOUL-Net obtain promising results in terms of both artifact suppression and detail preservation. The ROI indicated by the red boxes in the first row of \add{Fig. \ref{fig:xiongqiang}} is enlarged. It can be seen that the results of SOUL-Net maintain sharper edges and clearer structures, which can be better identified in the areas indicated by the red arrows and demonstrate the impressive potential in detail preservation of SOUL-Net. 
The SSIM and PSNR values of each reconstruction are given in the upper right of each subfigure of \add{Fig.\ref{fig:xiongqiang}}. Our proposed method outperforms all other methods in terms of PSNR and SSIM.

To better demonstrate the merits of SOUL-Net, \add{Fig. \ref{fig:abso_xiongqiang}} shows the absolute difference images associated with the reference image. It is obvious that our proposed SOUL-Net has minimal errors compared to other methods, which further confirms the effectiveness of our proposed method in noise and artifact suppression.

To further analyze the noise in the results of different methods in different frequency bands, the NPS maps are drawn in \add{Fig. \ref{fig:xiongqiang_NPS}}. It is clear that the proposed SOUL-Net outperforms all other methods in noise suppression.

\add{Fig. \ref{fig:xiongqiang_profile}} shows the horizontal profiles of different results along the blue line in the first row of \add{Fig. \ref{fig:xiongqiang}}. It can be seen that our proposed method obtains the closest results to the reference, especially near the edges, which are indicated by blue arrows.

\subsubsection{Quantitative Results}
\add{Tables \ref{tb:psnr_table}} and \add{\ref{tb:ssim_table}} show the average PSNR and SSIM across the whole test set in different energy bins. It is obvious that SOUL-Net achieves the best results in terms of both PSNR and SSIM in different energy bins, which demonstrates the effectiveness of our approach.

\subsection{Ablation experiments}
In this section, we conduct an experiment to verify the effectiveness of the attention mechanism. The quantitative results of SOUL-Net with and without an attention mechanism are shown in \add{Table \ref{tb:InitAbla}}. It can be concluded from the quantitative scores
that SOUL-Net outperforms SOUL-Net without the attention mechanism, which confirms the merit of the attention mechanism.
\begin{table}[!htb]
    \centering
    \setlength{\tabcolsep}{1.5mm}{
    \caption{AVERAGE PSNR VALUES OF RECONSTRUCTIONS WITH DIFFERENT ALGORITHMS ON TEST SET}
    \label{tb:psnr_table}
    \begin{tabular}{ccccccc} 
    \toprule
      Methods &  FBP & NLSMD  & LEARN & DSIR&ULTRA  & SOUL-Net \\
      \midrule
      60keV	&29.76	&38.52	&43.06&41.15&41.48&\textbf{44.84}\\
      70keV	&30.14&39.86	&43.25&42.47&42.35&\textbf{46.41}\\
      80keV&30.18&40.08	&44.20&42.60&42.39&\textbf{46.58}\\
      90keV&30.28&40.37	&44.81&43.04&42.50&\textbf{47.01}\\
      100keV&30.36&40.69&	45.27&43.21&42.75&\textbf{47.43}\\

      \bottomrule
      \end{tabular}}
\end{table}

\begin{table}[!htb]
    \centering
    \setlength{\tabcolsep}{1.5mm}{
    \caption{AVERAGE SSIM VALUES OF RECONSTRUCTIONS WITH DIFFERENT ALGORITHMS ON TEST SET}
    \label{tb:ssim_table}
    \begin{tabular}{ccccccc} 
    \toprule
      Methods &  FBP & NLSMD  & LEARN & DSIR&ULTRA  & SOUL-Net \\
      \midrule
      60keV	&0.5590&0.9431	&0.9757&0.9674&0.9712	&\textbf{0.9829}\\
      70keV	&0.5755&	0.9560	&0.9785&0.9754&	0.9773	&\textbf{0.9876}\\
      80keV&	0.5789	&0.9580	&0.9791&0.9757&	0.9776&\textbf{0.9873}\\
      90keV&	0.5826&	0.9602	&0.9819&0.9776&	0.9783	&\textbf{0.9886}\\
      100keV&	0.5875&	0.9633&	0.9831	&0.9794&0.9802&\textbf{0.9897}\\

      \bottomrule
      \end{tabular}}
\end{table}

\begin{table}[!t]
    \centering
    \setlength{\tabcolsep}{1.5mm}{
    \caption{QUANTITATIVE RESULTS WITH/WITHOUT ATTENTION}
    \label{tb:InitAbla}
    \begin{tabular}{ccc}
    \toprule
       Methods &SOUL-Net  &SOUL-Net without attention\\
      \midrule
      60keV &\textbf{44.84/0.9829}&44.57/0.9821\\
      70keV&\textbf{46.41/0.9876}&46.10/0.9869\\
      80keV&\textbf{46.58/0.9873}&46.30/0.9870\\
      90keV&\textbf{47.01/0.9886}&46.70/0.9881\\
      100keV&\textbf{47.43/0.9897}&47.12/0.9893\\
      \bottomrule
      \end{tabular}}
      
\end{table}

\section{DISCUSSION AND CONCLUSION}
In this paper, we proposed a
sparse and low-rank unrolling network for spectral CT image reconstruction. To make full use of the prior information of the spectral CT images, we introduced a learnable SVT method to learn low-rank regularization and used Taylor expansion-based method to calculate the backpropagation gradient. To better utilize the sparsity of spectral CT images, we used a convolutional neural network based on attention mechanism to implement sparse regularization. In the experimental results, our method demonstrated better qualitative and quantitative performance in both noise suppression and detail preservation than other state-of-the-art methods. This experimentally shows the effectiveness of our contributions.

Although the proposed SOUL-Net achieved promising results in spectral CT image reconstruction, it still has some drawbacks. One of the limitations is that the SVD operation in the SOUL-Net model requires additional time. Another limitation is the 
GPU memory consumption in SOUL-Net. If the data have more energy bins, then we need more GPU memory. The SVD operation also requires extra GPU memory. In addition, SOUL-Net requires paired data for training, which is usually difficult to obtain in clinical practice. Therefore in the future we will explore how to combine SOUL-Net with other unsupervised models, such as generative or flow models ~\cite{Noise2noise, GAN1, GAN2, GAN3, flow1, flow2}.


\begin{thebibliography}{00}

\bibitem{sct_intro_1}  Y. Long, \emph{et al}., “Multi-material decomposition using statistical image reconstruction for spectral CT,” \emph{IEEE Trans. Med. Imag.}, vol. 33, no. 8, pp. 1614-1626, 2014.
\bibitem{sct_intro_2} C. O. Schirra, \emph{et al}., “Statistical reconstruction of material decomposed data in spectral CT,” \emph{IEEE Trans. Med. Imag.}, vol. 32, no. 7, pp. 1249-1257, 2013.

\bibitem{sct_intro_3} L. Li, \emph{et al}., “Spectral CT modeling and reconstruction with hybrid detectors in dynamic-threshold-based counting and integrating models,” \emph{IEEE Trans. Med. Imag.}, vol. 34, no. 3, pp. 716-728, 2014.
\bibitem{intro_noise} P. M. Shikhaliev, “Projection X-ray imaging with photon energy weighting: Experimental evaluation with a prototype detector,” \emph{Phys. Med. Biol.}, vol. 54, no. 16, pp. 4971–4992, 2009.
\bibitem{art} Z. Long, \emph{\emph{et al}}. “A penalized Algebraic Reconstruction Technique (pART) for PET image reconstruction,” \emph{IEEE NSS}, Vol. 5. IEEE, 2007.

\bibitem{tv} E. Y. Sidky and X. Pan, “Image reconstruction in circular cone-beam computed tomography by constrained, total-variation minimization,” \emph{Phys. Med. Biol.}, vol. 53, no. 17, pp. 4777–4807, 2008.
\bibitem{tightFrame} B. Zhao, \emph{et al}. “Tight‐frame based iterative image reconstruction for spectral breast CT.” \emph{Med Phys.}, vol. 40, no.3, pp. 031905, 2013.

\bibitem{dictlearn} Q. Xu, \emph{et al}., “Low-dose X-ray CT reconstruction via dictionary learning,” \emph{IEEE Trans. Med. Imag.}, vol. 31, no. 9, pp. 1682–1697, Sep. 2012.


\bibitem{TDL} Y. Zhang, X. Mou, G. Wang, and H. Yu, “Tensor-based dictionary learning for spectral CT reconstruction,” \emph{IEEE Trans. Med. Imag.}, vol. 36, no. 1, pp. 142–154, Jan. 2017.

\bibitem{l0TDL} W. Wu, \emph{et al}., “Low-dose spectral CT reconstruction using image gradient $\ell_0$-norm and tensor dictionary,” \emph{Appl. Math. Model.}, vol. 63, pp. 538–557, 2018.
\bibitem{sct_spa} Y. Zhang, \emph{et al}., “Spectral CT reconstruction with image sparsity and spectral mean,” \emph{IEEE Trans. Comput. Imaging}, vol. 2, no. 4, pp. 510–523, Jan. 2016.

\bibitem{SISTER} D. Hu \emph{et al}. “SISTER: Spectral-image similarity-based tensor with enhanced-sparsity reconstruction for sparse-view multi-energy CT,” \emph{IEEE Trans. Comput. Imaging}, pp. 477-490, vol. 6 2019.

\bibitem{tensorNuclear} O. Semerci, \emph{et al}., “Tensor-based formulation and nuclear norm regularization for multienergy computed tomography,” \emph{IEEE Trans. Image Process.}, vol. 23, no. 4, pp. 1678–1693, Apr. 2014.

\bibitem{PRISM} H. Gao, H. Yu, S. Osher, and G. Wang, “Multi-energy CT based on a prior rank, intensity and sparsity model (PRISM),” \emph{Inverse Probl.}, vol. 27, no. 11, 2011, Art. no. 115012.
\bibitem{tPRISM} L. Li, \emph{et al}., “Spectral CT modeling and reconstruction with hybrid detectors in dynamic-threshold-based counting and integrating modes,” \emph{IEEE Trans. Med. Imag.}, vol. 34, no. 3, pp. 716–728, Mar. 2015.

\bibitem{SVT} J. F. Cai, E. J. Candès, and Z. Shen, “A singular value thresholding algorithm for matrix completion,” \emph{SIAM J. Optim.}, vol. 20, no. 4, pp. 1956-1982, 2008.
\bibitem{NLCTF} W. Wu, \emph{et al}. “Non-local low-rank cube-based tensor factorization for spectral CT reconstruction,” \emph{IEEE Trans. Med. Imaging}, vol. 38, no. 4, pp. 1079-1093, 2019.

\bibitem{LRTDSTTV} S. Li \emph{et al}. “An efficient iterative cerebral perfusion CT reconstruction via low-rank tensor decomposition with spatial–temporal total variation regularization,” \emph{IEEE Trans. Med. Imaging}, pp. 360-370, vol. 38, no. 2, 2018.

\bibitem{NLSMD} S. Niu, \emph{et al}., “Nonlocal low-rank and sparse matrix decomposition for spectral CT reconstruction,” \emph{Inverse Prob.}, vol. 34, no. 2, pp. 024003, 2018.

\bibitem{ASSIST} W. Xia, \emph{et al}., “Spectral CT reconstruction—ASSIST: aided by self-similarity in image-spectral tensors,” \emph{IEEE Trans. Comput. Imaging}, vol. 5, no. 3, pp. 420-436, 2019.

\bibitem{FONTSIR} X. Chen, \emph{et al}., “Convolutional sparse coding for compressed sensing CT reconstruction,” \emph{IEEE Trans. Med. Imaging}, 2022.

\bibitem{scss} B. Peng, \emph{et al}., “FONT-SIR: Fourth-Order Nonlocal Tensor Decomposition Model for Spectral CT Image Reconstruction,” \emph{IEEE Trans. Med. Imaging}, pp. 2607-2619, vol. 38, no. 11, 2019.









\bibitem{ImgDenoOv} C. Tian \emph{et al}. “Deep learning on image denoising: An overview,” \emph{IEEE Trans. Comput. Imaging}, pp. 251-275, vol. 131, 2020.

\bibitem{ImageSupSur} S. Minaee \emph{et al}. “Image segmentation using deep learning: A survey.” \emph{IEEE Trans. Pattern Anal. Mach. Intell.}, 2021.


\bibitem{ImgSupSur} Z. Wang \emph{et al}. “Deep learning for image super-resolution: A survey,” \emph{IEEE Trans. Pattern Anal. Mach. Intell.},vol. 43, no. 10, pp. 3365-3387, 2020.

\bibitem{CtSurvey1} C. M. McLeavy \emph{et al}. “The future of CT: deep learning reconstruction” \emph{Clin Radiol}, vol. 76, no. 6, pp. 407-415, 2021.

\bibitem{CtSurvey2} G. Wang \emph{et al}. “Deep learning for tomographic image reconstruction” \emph{Nat. Mach. Intell.}, vol. 2, no. 12, pp. 737-748, 2020.


\bibitem{CtSurvey3} E. Ahishakiye \emph{et al}. “A survey on deep learning in medical image reconstruction” \emph{Intell Medl}, vol. 1, no. 03, pp. 118-127, 2021.





\bibitem{ULTRA} W. Wu \emph{et al}. “Deep learning based spectral CT imaging,” \emph{Neural Netw}, pp. 342-358, no. 144, 2021.

\bibitem{DSIR} W. Mustafa \emph{et al}. “Sparse-View Spectral CT Reconstruction Using Deep Learning,” \emph{arXiv preprint}, 2020.
\bibitem{ADMM} J. Liu \emph{et al}. “Deep iterative reconstruction estimation (DIRE): approxi-mate iterative reconstruction estimation for low dose CT imaging,” \emph{Phys.Med. Biol.}, pp. 135007, vol. 64, 2019.

\bibitem{ISTA} A. Beck \emph{et al}. “A fast iterative shrinkage-thresholding algorithm for linear inverse problems,” \emph{SIAM J. Imaging Sci.}, pp. 183-202, vol. 2, no. 1, 2009.



\bibitem{RSVD} W. Wang \emph{et al}. “Robust differentiable svd,” \emph{IEEE Trans. Pattern Anal. Mach. Intell.}, 2021.

\bibitem{SVD_insta} C. Ionescu \emph{et al}. “Matrix backpropagation for deep networks with structured layers,” \emph{Proc. IEEE Conf. Comput. Vis.}, pp. 2965-2973, 2015.

\bibitem{MB} C. Ionescu \emph{et al}. “Matrix backpropagation for deep networks with structured layers.” \emph{Proc. Conf. Comput. Vis. Pattern Recognit.}, 2015.

\bibitem{MAGIC} W. Xia, \emph{et al}. “MAGIC: Manifold and graph integrative convolutional network for low-dose CT reconstruction,” \emph{IEEE Trans. Med. Imaging}, vol. 40, no. 12, pp. 3459-3472, 2021.

\bibitem{chambolle} A. Chambolle \emph{et al}. “An algorithm for total variation minimization and applications” \emph{J Math Imaging Vis}, vol. 20, no. 1, pp. 89-97, 2004.


\bibitem{SEnet} J. Hu \emph{et al}. “Squeeze-and-excitation networks” \emph{Proc. Conf. Comput. Vis. Pattern Recognit.}, pp. 7132-7141, 2018.

\bibitem{MedAtt1} D. Nie \emph{et al}. “ASDNet: attention based semi-supervised deep networks for medical image segmentation,” \emph{Proc. Int. Conf. Med. Image Comput. Comput.-Assisted Intervention}, pp. 370-378,2018.
\bibitem{MedAtt2} J. M. J.  Valanarasu \emph{et al}. “Medical transformer: Gated axial-attention for medical image segmentation,” \emph{Proc. Int. Conf. Med. Image Comput. Comput.-Assisted Intervention}, pp. 36-46, 2021.




\bibitem{DisDri} B. De. Man. \emph{et al}. “Distance-driven projection and backprojection in three dimensions,” \emph{Phys. Med. Biol.}, vol. 49, no. 11, pp. 2463, 2004.
\bibitem{AddNoise} S. Niu \emph{et al}. “Sparse-view X-ray CT reconstruction via total generalized variation regularization” \emph{Phys. Med. Biol.}, vol. 59, no. 12, pp. 2997–3017, 2014.

\bibitem{Adam} D. P. Kingma and J. Ba, “Adam: A method for stochastic optimization,”  \emph{arXiv preprint arXiv:}, 1412. 6980, 2014.

\bibitem{PSNRSSIM} W. Zhou, \emph{et al}., “Image quality assessment: from error visibility to structural similarity,” \emph{IEEE Trans. Image Process.}, vol. 13, no. 4, pp 600-612, 2004.
\bibitem{NPS} M. F. Kijewski, \emph{et al}., “The noise power spectrum of CT images” \emph{Phys. Med. Biol.}, vol. 32, no. 5, pp. 565, 1987.



\bibitem{NPS1} S. DOLLY, \emph{et al}., “Practical considerations for noise power spectra estimation for clinical CT scanners” \emph{Phys. Med. Biol.}, vol. 17, no. 3, pp. 392-407, 2016.

\bibitem{LEARN} H. Chen \emph{et al}. “LEARN: Learned experts’ assessment-based reconstruction network for sparse-data CT,” \emph{IEEE Trans. Med. Imag.}, pp. 1333-1347, vol. 37, no. 6, 2018.

\bibitem{Noise2noise} J. Lehtinen, \emph{et al}., “Noise2Noise: Learning image restoration without clean data” \emph{arXiv preprint arXiv}, 1803.04189, 2018.

\bibitem{GAN1} T. Karras, \emph{et al}., “A style-based generator architecture for generative adversarial networks” \emph{Proc. Conf. Comput. Vis. Pattern Recognit.}, 4401-4410, 2019.
\bibitem{GAN2} C. Tan, \emph{et al}., “A selective kernel-base cycle-consistent generative adversarial network for unpaired low-dose CT denoising” \emph{Precis. Clin. Med.}, vol. 5, no. 2, 2022.

\bibitem{GAN3} M. Ran, \emph{et al}., “Denoising of 3D magnetic resonance images using a residual encoder-decoder Wasserstein generative adversarial network” \emph{Med. Image Anal.}, pp. 165-180, vol. 55, 2019.
\bibitem{flow1} D. P. Kingma, \emph{et al}. “Glow: Generative flow with invertible 1x1 convolutions,” \emph{Proc. Adv. Neural Inf. Process. Syst.}, 2018, 31.
\bibitem{flow2} J. Ho, \emph{et al}. “Flow++: Improving flow-based generative models with variational dequantization and architecture design” \emph{Proc. Int. Conf. Mach. Learn.}, 2019, pp. 2722-2730.

\end{thebibliography}
\end{document}